\documentclass[12pt]{article}

\usepackage[comma,square,sort&compress]{natbib}
\usepackage{epsfig}
\usepackage{graphicx}

\DeclareMathAlphabet{\mathsc}{OT1}{cmr}{m}{sc}

\newcommand{\Sol}  {\mathsc{sol}}
\newcommand{\Atm}  {\mathsc{atm}}
\newcommand{\Sbl}  {\mathsc{sbl}}
\newcommand{\Nev}  {\mathsc{nev}}
\newcommand{\Lsnd} {\mathsc{lsnd}}
\newcommand{\dms}{\Delta m^2_\Sol}
\newcommand{\dma}{\Delta m^2_\Atm}
\newcommand{\dml}{\Delta m^2_\Lsnd}

\newcommand{\lesssim}{\:\mbox{\raisebox{-3pt}{$\stackrel%
{\displaystyle <}{\sim}$}}\:}
\newcommand{\gtrsim}{\:\mbox{\raisebox{-3pt}{$\stackrel%
{\displaystyle >}{\sim}$}}\:}

\begin{document}

\title{
\normalsize \hfill TUM-HEP-513/03\\[.5cm]
\large \bf Can four neutrinos explain global oscillation data
  including LSND \& cosmology?\footnote{Talk given by T.S.\ at the NOON 2003
workshop, February 10-14, 2003, Kanazawa, Japan}}
\author{M. Maltoni$^a$, T. Schwetz$^b$, 
  M.A. T\'ortola$^a$ and J.W.F. Valle$^a$ \\[2mm]
  \small $^a$ Instituto de F\'{\i}sica Corpuscular -- 
  C.S.I.C./Universitat de Val{\`e}ncia \\
  \small Edificio Institutos de Paterna, Apt 22085,
  E--46071 Valencia, Spain\\
  \small $^b$ Institut f{\"u}r Theoretische Physik, Physik Department\\
  \small Technische Universit{\"a}t M{\"u}nchen, 
  James-Franck-Str., D--85748 Garching, Germany}

\maketitle

\begin{abstract}
We present an analysis of the global neutrino oscillation data in
terms of four-neutrino mass schemes. We find that the strong preference of
oscillations into active neutrinos implied by solar+KamLAND as well as
atmospheric neutrino data allows to rule out (2+2) mass schemes, whereas (3+1)
schemes are strongly disfavoured by short-baseline experiments. In addition,
we perform an analysis using recent data from cosmology, including CMB data
from WMAP and data from 2dFGRS large scale structure surveys. These data lead
to further restrictions of the allowed regions for the (3+1) mass scheme.
\end{abstract}

\section{Introduction}

The neutrino oscillation interpretations of the solar\cite{solar,sno} and
KamLAND\cite{kamland} neutrino experiments, atmospheric\cite{skatm,macro}
neutrino data, and the LSND experiment\cite{lsnd} require three neutrino
mass-squared differences of different orders of
magnitude\cite{Pakvasa:2003zv}. Since it is not possible to obtain this within
the Standard Model framework of three active neutrinos it has been proposed to
introduce a light sterile neutrino\cite{sterile} to reconcile all the
experimental hints for neutrino oscillations. Here we present an analysis of
the global neutrino oscillation data in terms of four-neutrino mass schemes,
including data from solar, KamLAND and atmospheric neutrino experiments, the
LSND experiment, as well as data from short-baseline (SBL)
experiments\cite{KARMEN,CDHS,bugey} and reactor experiments\cite{lbl}
reporting no evidence for oscillations. We find that for all possible types of
four-neutrino schemes different sub-sets of the data are in serious
disagreement and hence, four-neutrino oscillations \textit{do not} provide a
satisfactory description of the global oscillation data including LSND. The
details of our calculations can be found in
Refs.~\cite{4nu01,4nu02,solat02}.

\section{Notations and approximations}

\begin{figure}[t]
 \centering
   \includegraphics[width=0.6\linewidth]{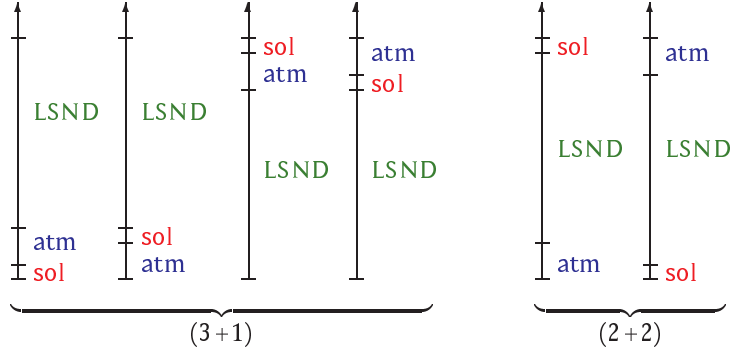}
      \caption{\label{fig:4spectra}%
        The six four-neutrino mass spectra, divided into the
        classes (3+1) and (2+2).}
\end{figure}

Four-neutrino mass schemes are usually divided into
the two classes (3+1) and (2+2), as illustrated in Fig.~\ref{fig:4spectra}. 
We note that (3+1) mass spectra include the three-active
neutrino scenario as limiting case. In this case solar and atmospheric
neutrino oscillations are explained by active neutrino oscillations,
with mass-squared differences $\dms$ and $\dma$, and the fourth
neutrino state gets completely decoupled. We will refer to such
limiting scenario as (3+0). In contrast, the (2+2) spectrum is
intrinsically different, as there must be a significant contribution
of the sterile neutrino either in solar or in atmospheric neutrino
oscillations or in both.

Neglecting CP violation, in general neutrino oscillations in
four-neutrino schemes are described by 9 parameters: 3 mass-squared
differences and 6 mixing angles in the unitary lepton mixing matrix.
Here we use a parameterisation introduced in Ref.~\cite{4nu01}, which
is based on physically relevant quantities: the 6 parameters $\dms$,
$\theta_\Sol$, $\dma$, $\theta_\Atm$, $\dml$, $\theta_\Lsnd$ are
similar to the two-neutrino mass-squared differences and mixing angles
and are directly related to the oscillations in solar, atmospheric and
the LSND experiments. For the remaining 3 parameters we use
$\eta_s,\eta_e$ and $d_\mu$. Here, $\eta_s \,(\eta_e)$ is the fraction
of $\nu_s \,(\nu_e)$ participating in solar oscillations, and
($1-d_\mu$) is the fraction of $\nu_\mu$ participating in oscillations
with $\dma$ (for exact definitions see Ref.~\cite{4nu01}). For the
analysis we adopt the following approximations:
(1) We make use of the hierarchy $\dms \ll \dma \ll \dml$.
This means that for each data set we consider only one mass-squared
difference, the other two are set either to zero or to infinity.
(2) In the analyses of solar and atmospheric data (but not for SBL data) we
set $\eta_e = 1$, which is justified because of strong constraints
from reactor experiments\cite{bugey,lbl}.

\begin{figure}[t]
 \centering
   \includegraphics[width=0.7\linewidth]{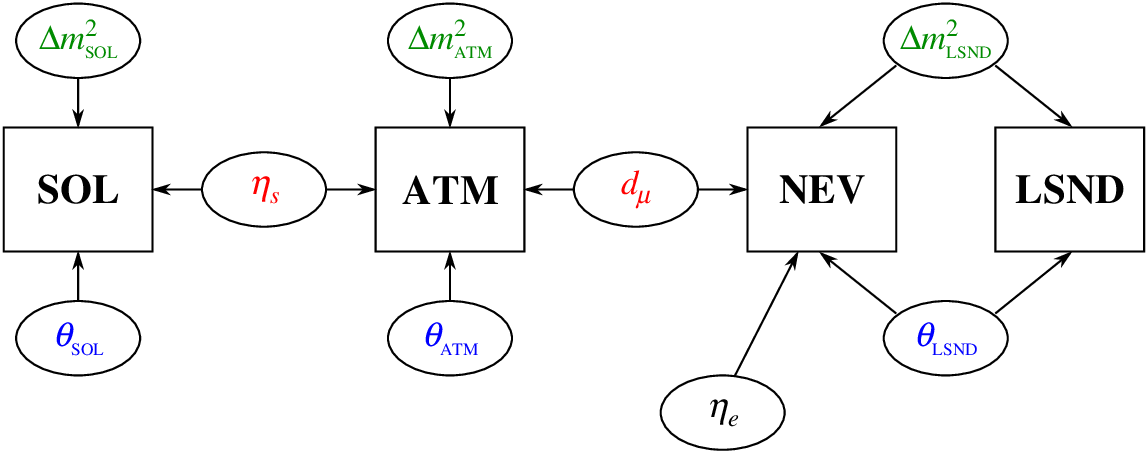}
      \caption{\label{fig:diagram}%
         Parameter dependence of the different data sets in our
         parameterisation. }
\end{figure}

Due to these approximations the parameter structure of the four-neutrino
analysis gets rather simple. The parameter dependence of the four data sets
solar, atmospheric, LSND and NEV is illustrated in Fig.~\ref{fig:diagram}. 
The NEV data set contains the experiments KARMEN\cite{KARMEN},
CDHS\cite{CDHS}, Bugey\cite{bugey} and CHOOZ/Palo Verde\cite{lbl}, reporting
no evidence for oscillations. We see that only $\eta_s$ links solar and
atmospheric data and $d_\mu$ links atmospheric and NEV data. LSND and NEV data
are coupled by $\dml$ and $\theta_\Lsnd$.

\section{(2+2): ruled out by solar and atmospheric data}
\label{sec:2+2}
The strong preference of oscillations into active neutrinos in solar and
atmospheric oscillations leads to a direct conflict in (2+2) oscillation
schemes. We will now show that thanks to the latest solar neutrino data
(especially from SNO\cite{sno}) in combination with the KamLAND
experiment\cite{kamland}, and the improved SK statistic on atmospheric
neutrinos\cite{skatm} the tension in the data has become so strong that (2+2)
oscillation schemes are essentially ruled out.\footnote{Details of our
analyses of the solar, KamLAND and atmospheric neutrino data can be found in
Refs.~\cite{solat02,KL}. For an earlier four-neutrino analysis of solar and
atmospheric data see Ref.~\cite{concha4nu}.} 

\begin{figure}[t]
  \centering
  \includegraphics[width=0.9\linewidth]{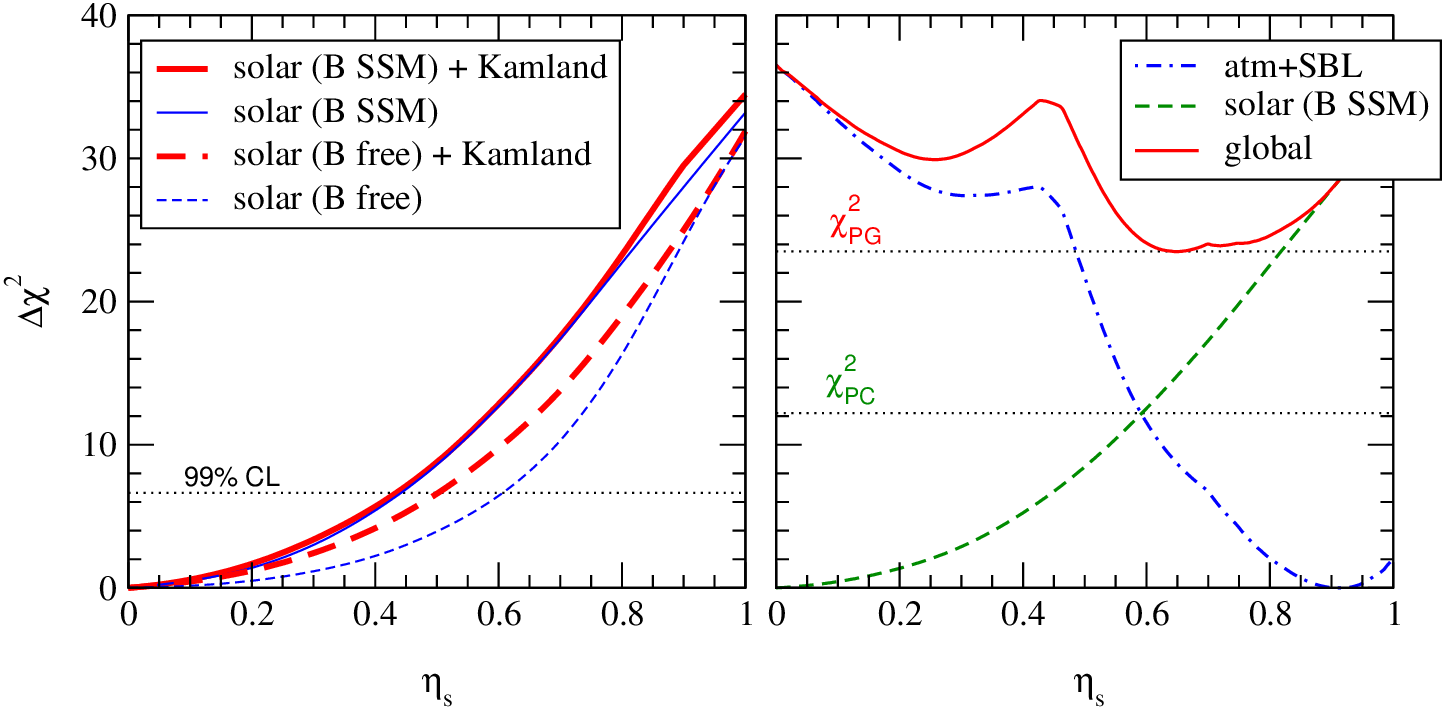}
  \caption{ Left: $\Delta\chi^2$ from solar and KamLAND data as a function of
    $\eta_s$. Right: $\Delta\chi^2_\Sol$, $\Delta\chi^2_{\Atm+\Sbl}$
    and $\bar\chi^2_\mathrm{global}$ as a function of $\eta_s$ in
    (2+2) oscillation schemes.}
  \label{fig:etas}
\end{figure}

The thin lines in the left panel of Fig.~\ref{fig:etas} show the
$\Delta \chi^2$ of latest solar data as a function of $\eta_s$, the
parameter describing the fraction of the sterile neutrino
participating in solar neutrino oscillations. The 99\% CL bounds
$\eta_s \le 0.44$, when the SSM constraint on the $^8$B-flux is
included, and $\eta_s \le 0.61$ for free $^8$B-flux reflect the strong
preference for active neutrino oscillations of solar data. Recently,
the outstanding results of the KamLAND reactor
experiment\cite{kamland} confirmed the LMA solution of the solar
neutrino problem\cite{KL,sol+KL}. Apart from this very important
result the sensitivity of KamLAND to an admixture of a sterile
neutrino is rather limited (see think lines in
Fig.~\ref{fig:etas}). The combined analysis leads to the 99\% CL
bound $\eta_s \le 0.5$ for $^8$B free, whereas in the SSM constrained
case the bound is unchanged.

In contrast, in (2+2) schemes atmospheric data imply $\eta_s \ge 0.65$
at 99\% CL, in clear disagreement with the bound from solar data.  In
the right panel of Fig.~\ref{fig:etas} we show the $\chi^2$ for solar
data and for atmospheric combined with SBL data as a function of
$\eta_s$.  Furthermore, we show the $\chi^2$ of the global data
defined by
\begin{equation}\label{chi2solatm}
\bar\chi^2(\eta_s) \equiv 
\Delta\chi^2_\Sol(\eta_s) + 
\Delta\chi^2_{\Atm + \Sbl}(\eta_s) \,.
\end{equation}
From the figure we find that only if we take both data sets at the 99.95\% CL
a value of $\eta_s$ exists, which is contained in the allowed regions of both
sets. This follows from the $\chi^2$-value $\chi^2_\mathrm{PC} = 12.2$ shown
in the figure. In Refs.~\cite{4nu02,PG} we have proposed a statistical
method to evaluate the disagreement of different data sets in global analyses.
The \textit{parameter goodness of fit} (PG) makes use of the $\bar\chi^2$
defined in Eq.~(\ref{chi2solatm}). This criterion evaluates the GOF of the
\textit{combination} of data sets, without being diluted by a large number of
data points, as it happens for the usual GOF criterion (for details see
Ref.~\cite{PG}). We find $\chi^2_\mathrm{PG} \equiv \bar\chi^2_\mathrm{min}
= 23.5$, leading to the marginal PG of $1.3 \times 10^{-6}$.  We conclude that
(2+2) oscillation schemes are highly disfavoured by the disagreement between
the latest solar and atmospheric neutrino data. This is a very robust result,
independent of whether LSND is confirmed or disproved.\footnote{Sub-leading
effects beyond the approximations adopted here should not affect this result
significantly. Allowing for additional parameters to vary might change the
{\it ratio} of some observables\cite{Pas:2002ah}, however, we expect that the
absolute number of events relevant for the fit will not change substantially.}

\section{(3+1): strongly disfavoured by SBL data}

\begin{figure}[t]
  \centering 
  \includegraphics[width=0.7\linewidth]{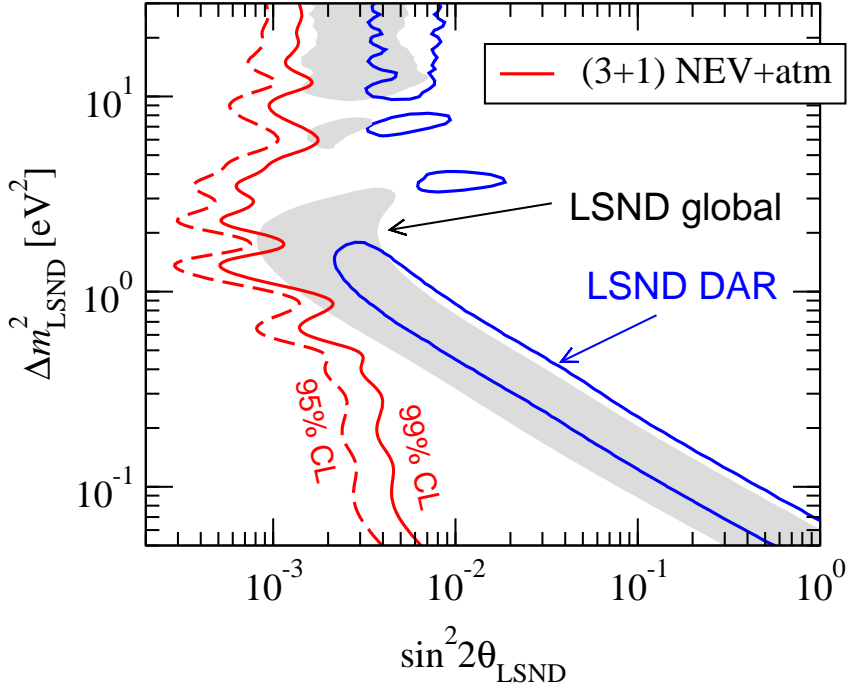}
  \caption{Upper bound on $\sin^22\theta_\Lsnd$ from NEV and
  atmospheric neutrino data in (3+1) schemes\protect\cite{cornering}
  compared to the allowed region from global LSND
  data\protect\cite{lsnd} and decay-at-rest (DAR) LSND
  data\protect\cite{Church:2002tc}.}
  \label{fig:3+1}
\end{figure}

It is known for a long time\cite{3+1early,okada} that (3+1) mass schemes are
disfavoured by the comparison of SBL disappearance
data\cite{CDHS,bugey} with the LSND result. In
Ref.~\cite{cornering} we have calculated an upper bound on the LSND
oscillation amplitude $\sin^22\theta_\Lsnd$ resulting from NEV and
atmospheric neutrino data. From Fig.~\ref{fig:3+1} we see that this
bound is incompatible with the signal observed in LSND at the 95\%
CL. Only marginal overlap regions exist between the bound and global
LSND data if both are taken at 99\% CL. An analysis in terms of the
parameter goodness of fit\cite{4nu02} shows that for most values of
$\dml$ NEV and atmospheric data are compatible with LSND only at more
than $3\sigma$, with one exception around $\dml \sim 6$ eV$^2$, where
the PG reaches 1\%. These results show that (3+1) schemes are strongly
disfavoured by SBL disappearance data.

\section{Comparing (3+1), (2+2) and (3+0) hypotheses}

With the methods developed in Ref.~\cite{4nu01} we are able to
perform a global fit to the oscillation data in the four-neutrino
framework. This approach allows to statistically compare the different
hypotheses. Let us first evaluate the GOF of (3+1) and (2+2) spectra
with the help of the PG method described in Ref.~\cite{PG}. We
divide the global oscillation data in the 4 data sets SOL, ATM, LSND
and NEV. Following Ref.~\cite{4nu02}
we consider
\begin{equation}\label{chi2bar}
\begin{array}{ccl}
\bar\chi^2 &=&
\Delta\chi^2_\Sol(\theta_\Sol,\dms,\eta_s) 
+ \Delta\chi^2_\Atm(\theta_\Atm,\dma,\eta_s,d_\mu) \\
&+& \Delta\chi^2_\Nev(\theta_\Lsnd,\dml,d_\mu,\eta_e) 
+ \Delta\chi^2_\Lsnd(\theta_\Lsnd,\dml) \,,  
\end{array}
\end{equation}
where $\Delta\chi^2_X = \chi^2_X - (\chi^2_X)_\mathrm{min}$ ($X$ =
SOL, ATM, NEV, LSND). In Tab.~\ref{tab:pg} we show the contributions of
the 4 data sets to $\chi^2_\mathrm{PG} \equiv \bar\chi^2_\mathrm{min}$
for (3+1) and (2+2) oscillation schemes. As expected we observe that
in (3+1) schemes the main contribution comes from SBL data due to the
tension between LSND and NEV data in these schemes. For (2+2)
oscillation schemes a large part of $\chi^2_\mathrm{PG}$ comes from
solar and atmospheric data, however, also SBL data contributes
significantly.  This comes mainly from the tension between LSND and
KARMEN\cite{Church:2002tc}, which does not depend on the mass scheme
and, hence, also contributes in the case of (2+2). Therefore, the
values of $\chi^2_\mathrm{PG}$ in Tab.~\ref{tab:pg} for (2+2) schemes
are higher than the one given in Sec.~\ref{sec:2+2}, where the tension
in SBL data is not included.

\begin{table}[t]\centering
    \catcode`?=\active \def?{\hphantom{0}}     
    \begin{tabular}{|c|cccc|c|c|}
    \hline
    & SOL & ATM & LSND & NEV &   $\chi^2_\mathrm{PG}$ & PG \\
    \hline 
(3+1) & ?0.0 & 0.4 & 7.2 & 7.0 & 14.6 & $5.6 \times 10^{-3}$ \\
(2+2) & 14.8 & 6.7 & 2.2 & 9.7 & 32.4 & $1.6 \times 10^{-6}$ \\
    \hline
    \end{tabular}
    \caption{Parameter GOF and the contributions of different data sets to 
    $\chi^2_\mathrm{PG}$ in (3+1) and (2+2) neutrino mass schemes.}
    \label{tab:pg}
\end{table}

The parameter goodness of fit is now obtained by evaluating
$\chi^2_\mathrm{PG}$ for 4 DOF\cite{PG}. This number of degrees of
freedom corresponds to the 4 parameters $\eta_s, d_\mu, \theta_\Lsnd,
\dml$ describing the coupling of the different data sets (see
Eq.~(\ref{chi2bar}) and Fig.~\ref{fig:diagram}). The best GOF is
obtained in the (3+1) case. However, even in this best case the PG is
only 0.56\%. The PG of $1.6\times 10^{-6}$ for (2+2) schemes shows
that these mass schemes are essentially ruled out by the disagreement
between the individual data sets.

Although we have seen that none of the four-neutrino mass schemes can provide
a reasonable good fit to the global oscillation data including LSND, it might
be interesting to consider the \textit{relative} status of the three
hypotheses (3+1), (2+2) and the three-active neutrino scenario (3+0). This can
be done by comparing the $\chi^2$ values of the best fit point (which is in
the (3+1) scheme) with the one corresponding to (2+2) and (3+0).  First we
observe that (2+2) schemes are strongly disfavoured with respect to (3+1) with
a $\Delta \chi^2 = 17.8$, implying a CL of 99.87\% for 4 DOF. Further we find
that (2+2) is only slightly better than (3+0), which is disfavoured with a
$\Delta \chi^2 = 20.0$ with respect to (3+1).

\section{Adding information from cosmology}

Recently a precision measurements of the cosmic microwave background (CMB)
radiation has been published by the WMAP collaboration\cite{wmap}. Under
certain assumptions about the cosmological model and in combination with other
cosmological data a stringent upper bound on the sum of the neutrino masses
$\sum m_\nu \lesssim 0.7$ eV is obtained\cite{wmap,Elgaroy:2003yh}. Since this
bound is of the same order as $\sqrt{\dml}$ it will lead to further
restrictions in four-neutrino mass schemes\cite{wmap3+1}. To combine this
information with data from oscillation experiments we use the results of an
analysis performed by S.~Hannestad\cite{hannestad}. There, WMAP\cite{wmap}
results have been combined with other CMB data\cite{CMB}, data from the large
scale structure survey 2dFGRS\cite{LSS}, the HST Hubble key project\cite{HST},
and Type Ia supernovae observations\cite{SN}. It has been found that a
non-zero $\sum m_\nu$ can be compensated partially by an increase in the
effective number of neutrino species. Therefore, the original bound obtained
in Ref.~\cite{wmap}, which was calculated assuming three neutrinos, cannot
be applied in a straightforward way. In the case of four neutrinos a somewhat
weaker bound of $\sum m_\nu < 1.38$ eV (95\% CL) is obtained\cite{hannestad}.
To combine this result with data from oscillation experiments we use the
$\chi^2$ from cosmology for four neutrinos, shown in Fig.~6 of
Ref.~\cite{hannestad}.

\begin{figure}[t]
  \centering \includegraphics[width=0.9\linewidth]{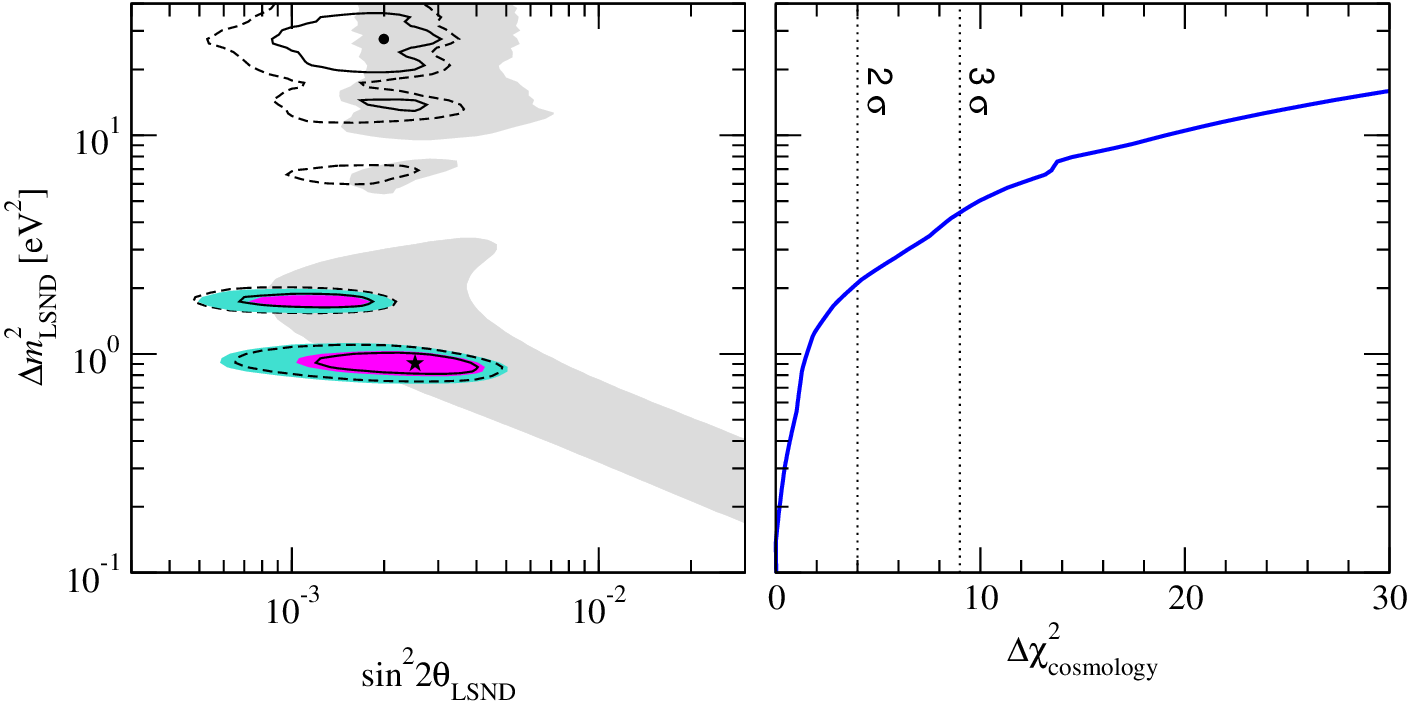}
\caption{Left: Allowed regions at 90\% and 99\% CL for (3+1) schemes without
(solid and dashed lines) and including data from cosmology (coloured regions).
The grey region is the 99\% CL region of LSND\protect\cite{lsnd}. Right:
$\Delta\chi^2$ from cosmological data\protect\cite{hannestad} as a function of
$\dml$.} \label{fig:cosmology}
\end{figure}

In the following analysis we consider (3+1) schemes of the hierarchical type
(first two spectra in Fig.~\ref{fig:4spectra}) and neglect the masses of the
three lower mass states, such that $\sum m_\nu \approx
\sqrt{\dml}$.\footnote{Note that all other four neutrino schemes will be much
more disfavoured, since more mass states will contribute at the LSND mass
scale.} The $\Delta \chi^2_\mathrm{cosmology}$ from Ref.~\cite{hannestad}
under this assumption is shown in the right panel of Fig.~\ref{fig:cosmology}
as a function of $\dml$. From the left panel of this figure one finds that
cosmology excludes the region $\dml \gtrsim 3$ eV$^2$. However, the islands at
$\dml \sim 0.91$ eV$^2$ and 1.7 eV$^2$ are practically unaffected by
cosmological data. Performing a GOF analysis shows that adding cosmological
data gives $\chi^2_\mathrm{PG} = 17.1$ and 19.6 for the two islands. This
should be evaluated for 5 DOF\cite{PG}, leading to a PG of 0.43\% and 0.15\%,
respectively, which is comparable to the value given for (3+1) in
Tab.~\ref{tab:pg}. We note, however, that a modest improvement of the bound on
$\sum m_\nu$ by future cosmological data will drastically worsen the fit of
the last remaining islands in the four-neutrino parameter space. 

A further cosmological constraint on four-neutrino schemes comes from Big Bang
nucleo synthesis (BBN)\cite{okada,BGGS}. In the absence of any non-standard
mechanism, like a large lepton number asymmetry, (3+1) as well as (2+2)
schemes will lead to an effective number of neutrino species $N_\nu =
4$\cite{BBNeq}. This is in conflict with the upper bound on $N_\nu$ derived 
in recent analyses\cite{hannestad,Crotty:2003th,Barger:2003zg} including CMB
and BBN data. Taking into account the uncertainty related to the primordial
abundances of He and D, upper bounds in the range $N_\nu < 3.1-3.3$
($2\sigma$)\cite{Barger:2003zg} are obtained.

\section{Conclusions}
Performing a global analysis of current neutrino oscillation data we find that
four-neutrino schemes do not provide a satisfactory fit to the data:

$\bullet$ 
%\begin{itemize}
%\item
The strong rejection of non-active oscillation in the solar+KamLAND
and atmospheric neutrino data rules out (2+2) schemes, independent of whether
LSND is confirmed or not. Using an improved goodness of fit method especially
sensitive to the combination of data sets we obtain a GOF of only $1.6\times
10^{-6}$ for (2+2) schemes.

$\bullet$ 
%\item 
(3+1) spectra are disfavoured by the disagreement of LSND with
short-baseline disappearance data, leading to a marginal GOF of $5.6\times
10^{-3}$.  If LSND should be confirmed we need more data on $\nu_e$ and/or
$\nu_\mu$ SBL disappearance to decide about the status of (3+1).

$\bullet$ 
%\item 
Recent cosmological data lead to further restrictions of the
allowed parameter space. Only two islands aground $\dml \sim 0.91$ eV$^2$ and
1.7 eV$^2$ in hierarchical (3+1) schemes survive.
%\end{itemize}

Our analysis brings the LSND hint to a more puzzling status, and the situation
will become even more puzzling if LSND should be confirmed by the up-coming
MiniBooNE experiment\cite{miniboone}. Recently an explanation in terms of five
neutrinos has been proposed\cite{Sorel:2003hf}. In such a (3+2) scheme solar
and atmospheric data are explained dominantly by active oscillations, similar
to the (3+1) case. The fit of SBL data is improved with respect to (3+1) by
involving two large mass splittings $\Delta m^2_{41} \sim 0.9$ eV$^2$ and
$\Delta m^2_{51} \sim 20$ eV$^2$. Alternative explanations of LSND
by anomalous muon decay\cite{Babu:2002ic} seem to be in disagreement
with the negative result of KARMEN, or require a very radical modification of
standard physics, like CPT violation\cite{Barenboim:2002ah}.

{\bf Acknowledgements:}
T.S.\ would like to thank the organisers for the very interesting workshop and
for financial support. Furthermore, we thank S.~Hannestad for providing us
with the $\chi^2$ of cosmological data, and S.~Pastor for discussions.  This
work was supported by Spanish grant BFM2002-00345, by the European Commission
RTN grant HPRN-CT-2000-00148 and by the ESF Neutrino Astrophysics Network. 
M.M.\ was supported by Marie Curie contract HPMF-CT-2000-01008, T.S.\ was
supported by the ``Sonderforschungsbereich 375-95 f{\"u}r
Astro-Teilchenphysik'' der Deutschen Forschungsgemeinschaft and M.A.T.\ was
supported by the M.E.C.D.\ fellowship AP2000-1953.

\end{document}